\documentclass[preprints,article,accept,pdftex,moreauthors]{my-mdpi} 

\firstpage{1} 
\pubvolume{11}
\issuenum{9}
\articlenumber{457}
\pubyear{2022}
\copyrightyear{2022}
\externaleditor{Academic Editor: Joao Paulo Carvalho and Oscar Humberto Montiel Ross} 
\datereceived{05 July 2022} 
\dateaccepted{01 September 2022} 
\datepublished{06 September 2022} 
\hreflink{https://doi.org/10.3390/axioms11090457} 
\pdfoutput=1

\Title{A Schelling Extended Model in Networks{---} Characterization of Ghettos in Washington D.C.}

\TitleCitation{A Schelling Extended Model in Networks {---}Characterization of Ghettos in Washington D.C.}

\Author{Diego Ortega $^{1,}$*\orcidA{} and Elka Korutcheva $^{1,2}$\orcidB}

\AuthorNames{Diego Ortega and Elka Korutcheva}

\AuthorCitation{Ortega, D.; Korutcheva, E.}

\address{%
$^{1}$ \quad Dto. Física Fundamental, Universidad Nacional de Educación a Distancia (UNED), E-28040 Madrid,  Spain 
\\
$^{2}$ \quad G. Nadjakov Institute of Solid State Physics, Bulgarian Academy of Sciences, 1784 Sofia, Bulgaria}

\corres{Correspondence: dortega144@alumno.uned.es}

\abstract{Segregation affects millions of urban dwellers. The main expression of this reality is the creation of \textit{ghettos} which are city parts characterized by a combination of features: low income, poor cultural level... Segregation models have been usually defined over regular lattices. However, in recent years, the focus has shifted from these unrealistic frameworks to other environments defined via geographic information systems (GIS) or networks. Nevertheless, each one of them has its drawbacks:  GIS demands high-resolution data, that are not always available, and networks tend to have limited real-world applications. Our work tries to fill the gap between them.  First, we use some basic GIS information to define the network, and then, run an extended Schelling model on it.  As a result, we obtain the location of ghettos. After that, we analyze which parts of the city are segregated, via spatial analysis and machine learning and compare our results. For the case study of Washington D.C., we obtain an $80\%$ accuracy.}


\begin{document}

\section{Introduction}

A ghetto is a part of a city in which members of a minority group live, especially as a result of political, social, legal, environmental, or economic pressure \cite{Merriam-Webster(2022)}. Regrettably, millions of people live in ghettos, bringing out the magnitude of the matter. Although their common feature is the impoverishment of the zone, different kinds of ghettos can be found across the world.  

Ghettos constitute the main expression of segregation. Segregation can be understood as the practice of separating groups of people with differing characteristics, often connoting a condition of inequality \cite{Britannica(2022)}. Even though segregation can have its origin in economic, cultural, or religious motives, we focus on the racial one. Racial segregation restricts people of a different race to areas whose facilities have lower standards than the rest of the city, i.e., ghettos. An actual example of these is some black neighborhoods in the city of Washington D.C., our case study.

The understanding and measuring of segregation can be complex. On one hand, models of segregation have been studied from the sociophysics field \cite{Schelling(1971), Blume(1971), Ortega(2021),Vinkovic(2006), Dallasta(2008), Gauvin(2010), Urselmans(2018),Zhang(2004), Fosset(2006), Jensen(2018), Flaig(2019)}. Whereas these contributions provide deep insights into different aspects of segregation, they take place in simple square lattices. On the other hand,  GIS allows us to define more realistic frameworks. However, the measure of segregation itself has drawn substantial attention during the last thirty years \cite{Yao(2018), Anselin(1995), Duncan(1955), Massey(1988), osullivan(2007), Farber(2015), Harris(2017), Reitano(2020)}, but a consistent and generally agreeable definition of segregation has not been formulated yet. The ultimate trend in the field consists in the use of microdata to minimize the modifiable areal unit problem (MAUP) \cite{Jelinski(1996)}. Nevertheless, this approach requires the use of data on the personal level which is not always available.

In this article, we propose a method to bridge the gap between Sociophysics models and the measure of segregation by GIS techniques. Our procedure relies on capturing the ghettos' location with a segregation model and basic data from the city in situations where detailed data can be not completely reliable. This is a novel approach to the segregation field, as we can see in Section~\ref{RW}. We start by using basic GIS information from our region of interest to define a network and then, run on this framework an extended Schelling \mbox{model \cite{Schelling(1971)}.} Our enhanced version takes into account the economic contribution to segregation, including terms for the housing market and the financial gap between the population. The final result is the location of ghettos on the network, which can be mapped into their corresponding city areas. Finally, these predicted ostracized regions are compared to the ones characterized as ghettos by spatial analysis 
 (SA) and Machine Learning (ML) algorithms. For the real case study of Washington D.C., the obtained accuracy is $80 \pm 7 \%$.

\begin{figure}[H]
\begin{adjustwidth}{-5cm}{-1.5cm}
\centering
\includegraphics[width=17.5 cm, height=2.3cm]{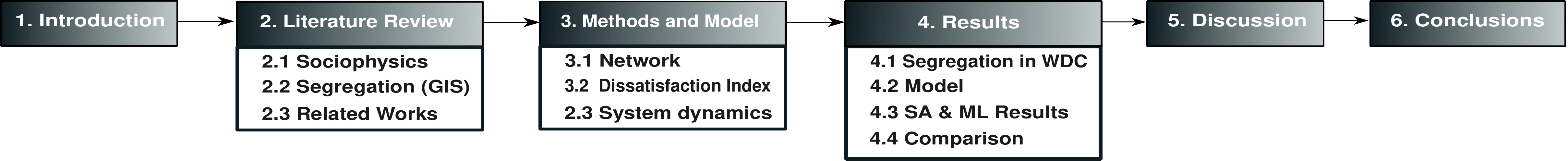}
\end{adjustwidth}
\caption{Scheme of the paper organization. WDC, SA, and ML are the abbreviations of Washington D.C, Spatial Analysis, and Machine Learning, respectively. }
\label{fig:1}
\end{figure}

The paper is organized as follows. In Section \ref{lr} the segregation literature previously cited is reviewed. Section \ref{model} explains how to define the network and discusses the system dynamics. In Section \ref{result} we describe our findings, accentuating the connection between the simple network model and the real segregation phenomena. Discussion of our results is found in Section~\ref{disc}. Finally, our conclusions and proposals for further work are explored in Section~\ref{conc}. A guideline is illustrated in Figure \ref{fig:1}.

\section{Literature Review}
\label{lr}

As our work involves the study of segregation from the Sociophsyics and Geographic fields we try to cover their evolution in Sections~\ref{SP} and \ref {SG}, respectively. Papers related to our work are discussed in Section~\ref{RW}. 

\subsection{The Schelling Model and Other Related Models from Sociophysics}
\label{SP}

One of the first approaches to the segregation field was put forward by T. C. \linebreak Schelling \cite{Schelling(1971)}. His work considered two different social groups (\textit{red }and \textit{blues}) distributed over a square lattice with some vacancies. As people tend to seek out neighbors who are similar to themselves, large clusters of the same type of agents are typically created. Each agent has an \textit{state} and an \textit{action}. An agent is in a state of happiness when the fraction of different neighbors in her/his neighborhood, $f_d$, does not exceed the tolerance value $T$. However, if $f_d > T$, the agent is unhappy. Regarding actions, an unhappy agent can relocate in a vacancy if happiness can be attained in the new location. Otherwise, the agent is characterized as happy and no action is performed. Several extensions of this model have been developed and can be classified into two broad categories: more physically oriented or socioeconomically centered.

The Schelling model is related to the Blume-Emery-Griffiths model (BEG), which was originally used to study $He^3$-$He^4$ mixtures \cite{Blume(1971)}. This framework can be linked to the Schelling model by associating the type of agents with the spin values considered in their work: {\em red} and {\em blue} agents with $s=\pm1$, respectively, while vacancies imply a null spin. For certain parameter values, the Hamiltonian of the system is reduced when similar agents group together and clusters of different colors are separated by vacancies \cite{Ortega(2021)}. In addition to this, the number of vacancies is controlled by a crystal field which plays the role of a chemical potential. The incorporation of an external magnetic field favors one spin type over the other one, giving rise to dissimilar entry fluxes for each agent type.

A physical analog of the model comparing segregation between clusters with two liquids was characterized in \cite{Vinkovic(2006)}, an approach from the statistics physics to the segregation phenomena was carried out in \cite{Dallasta(2008)}, a relation with a 1-spin model where agents can enter or leave the city (open city) was established in \cite{Gauvin(2010)}. From the social and economic fields, a housing market and different tolerance values for black and white people were introduced in the simulation \cite{Zhang(2004)}. Another interesting contribution was able to predict the relocation of higher-status households in suburban zones by using a cellular-automata  \cite{Fosset(2006)}. 

In recent years focus has shifted to complex agent-based models which include ethical issues: the influence of altruistic agents which greatly affects the final state of the system improving the overall happiness \cite{Jensen(2018)}. The influence of altruist and fair agents is evaluated in \cite{Flaig(2019)}, the behavior of agents with adaptive tolerance to their neighborhood in  \cite{Urselmans(2018)} and open city simulations, in which agents can leave or enter the city, were used to model gentrification processes in \cite{Ortega(2021)}. 

Nevertheless, these models consider the classical rectangular lattice whose applications to urban realities are limited in most cases. The GIS technology solved this issue by mapping real cities and measuring segregation in them.

\subsection{Segregation (GIS)}
\label{SG}
Despite being a reality in urban environments, the measurement of segregation is a complex problem that emerges as a special spatial arrangement across multiple dimensions of culture, religion, economy, race, and others \cite{Yao(2018)}. The clustering of these groups in parts of the city gives rise to the creation of ghettos. Five dimensions of segregation: evenness, exposure-isolation, concentration, centralization and clustering were proposed in \cite{Massey(1988)}. It seems that these variables overlap, simplifying the measure of segregation to one index in most cases. One of the most used estimators is the dissimilarity index \cite{Duncan(1955)}, which compares how evenly one population sub-group is spread out geographically compared to another population sub-group. However, this index does not take advantage of all the spatial information \cite{Yao(2018)}. Therefore, segregation estimators taking advantage of the clustered nature of the phenomena were developed. Although several local indicators of spatial association (LISA) statistics were proposed \cite{osullivan(2007)}, the local Moran´s indices \cite{Anselin(1995)} remain one of the most used.

Nevertheless, all the measures previously cited rely on data obtained from groups of individuals. Hence, these measures are exposed to the modifiable areal unit problem (MAUP) \cite{Fotheringham(1991), Jelinski(1996)}. This issue is created by the modification of the boundaries in geographical units which are often demarcated artificially, i.e., they are not natural divisions. Therefore new measures relying on personal patterns which characterize people from the same neighborhood were proposed  \cite{Farber(2015)}. Nowadays, segregation is usually considered a multiscale phenomenon, and microdata is used when available \cite{Benenson(2009),  Harris(2017), Reitano(2020)}. 

Although an approach based on the minimization of the MAUP requires an individual level analysis, it also makes us dependent on the microdata level which is not always available or can be affected by variations such as the one provoked by the covid-19 pandemic.

\subsection{Related Works}
\label{RW}

Works particularly related to our contribution are those using the Schelling model on networks. The major role played by mild tolerance preferences in the segregation phenomena was compared for lattice and networks in \cite{Fagiolo(2007)}. This work also explained that polarization mechanisms occur not only in regular spatial networks but also in more general social networks. The performance of some segregation indices is particularized for several network types in \cite{Cortez(2015)}. Besides, the dynamics of the model are also characterized, thus finding that the system evolves toward steady states in which a maximum level of segregation is reached. In contrast to the previous works, where similar results are found despite the inherent differences between networks, the importance of cliques is underlined in \cite{Banos(2010)}. Cliques are complete subgraphs inside another graph. They can be understood as clusters that reinforce segregation effects. Hence, suggestions against real city situations that can generate structures similar to cliques are put forward.

None of the previous network models have a direct connection with an urban structure, in contrast to the GIS-related works. In \cite{Crooks(2019)}, several examples of the interaction between GIS and agent-based models are presented. One of them is specially related to this work, given that is a segregation model in which each census tract represents an agent in the Schelling model. A segregation model loosely coupled with GIS information is proposed \mbox{in \cite{Crooks(2009)}.} In this contribution, several agents can occupy the same polygon. In addition to this, the significance of barriers in segregation is evaluated. Basically, borders, such as rivers and highways, tend to amplify this urban reality. This finding is in good agreement with the significance of cliques in networks previously discussed \cite{Banos(2010)}.

The models we have briefly explained in this section assume a perspective of monetary equality for the agents, i.e., no economic gap between the groups. In addition to this, no housing price is associated with city areas. Therefore, they lead to a portrayal focused on clusterization where no identification of the economically handicapped group is possible. To put it in other words, ghettos can not be located by adopting these frameworks.

\section{Methods and Model}
\label{model}

\subsection{Network}
\label{GN}

\begin{figure}[H]
\begin{tabular}{cc}
\includegraphics[width=6.00cm, height=4.0cm]{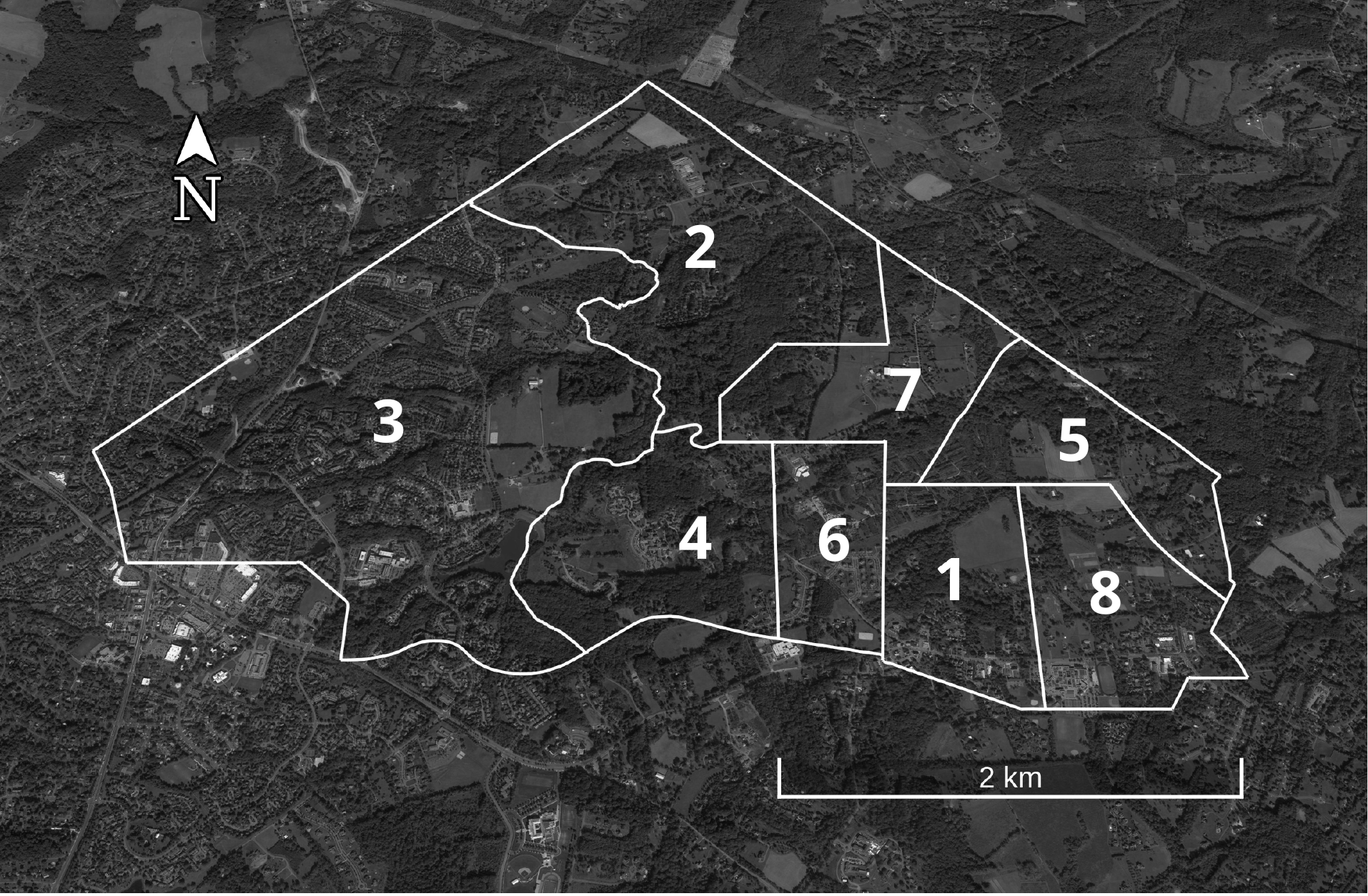}&
\includegraphics[width=6.00cm, height=4.0cm]{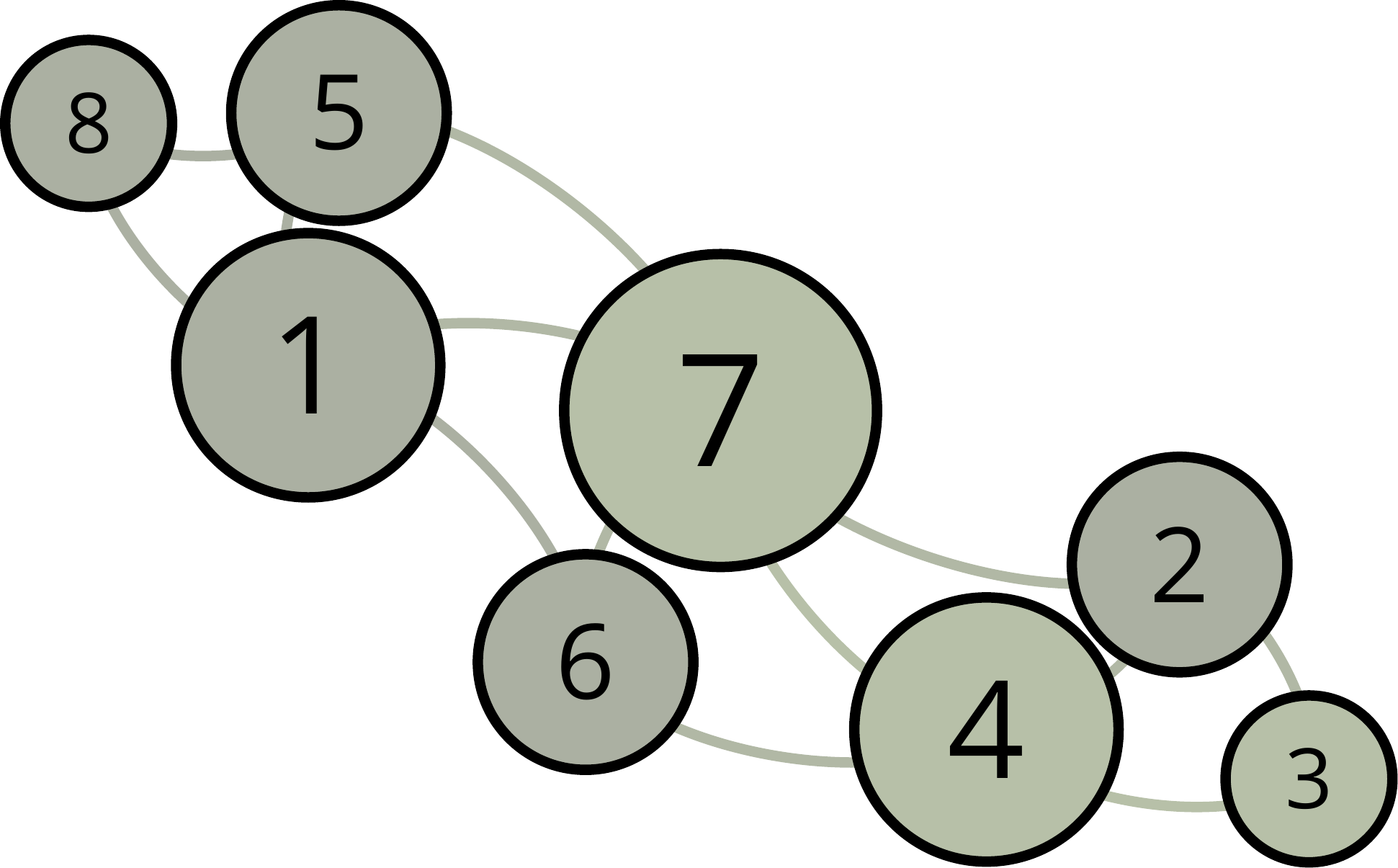}\\
(\textbf{a})&(\textbf{b})\\
\end{tabular}
\caption{(\textbf{a}) Our region is divided in different polygons by census tracts. Each polygon is numbered to facilitate its identification. (\textbf{b}) Each node corresponds to a polygon with the same identification number as the figure on the left. Edges link overlapping polygons from the previous illustration. Node sizes are related to their degree.}
\label{fig:2}
\end{figure}

In this subsection, we explain the process of building our network. We consider a small zone in the north of Washington D.C. for pedagogical purposes. First, we need access to this census tract data. Their boundaries divide the region into different polygons, as can be seen in Figure \ref{fig:2}a. It is worth noting that information on the different regions, associated with geographical, economic, or racial aspects could be linked to each node.

Our network is defined by the nodes that represent each polygon from the map, and edges are established between neighbor polygons, as it is shown in Figure \ref{fig:2}b. As a means to obtain these relations, QGIS software is applied \cite{qgis(2022)}. Then, this data is mapped into a network by using the networkx package \cite{networkx(2008)}.

\subsection{Dissatisfaction Index}
\label{DI}
Once our network is established we focus our attention on the agents. As in the Schelling model, we consider two kinds of agents defined by its color: red and blue. Each agent occupies a node, while some nodes may remain empty. 

The fraction of different neighbors for the agent in the node \textit{i} can be written as  $f_d(i)={N_d}/{(N_s+N_d)}$, where $N_s$ and $N_d$ are, respectively, the number of similar (s) and dissimilar (d) agents in the neighborhood, i.e., the linked nodes. In contrast to lattice models, where the numbers of neighbors remain fixed, the number of neighbors can vary from one node to another in our network.  

The state of the agent in the node \textit{i} is measured via the dissatisfaction index, $I_{dis}(i)$ as:

\begin{equation}
  I_{dis}(i)=f_{d}(i) - T+ D(i) \pm H.
  \label{eq:1}
\end{equation}

Here,  the first two terms are related to social preferences. $T$ is the tolerance, the maximum fraction of different agents that an agent can withstand while remaining happy. This value acts as a threshold. These two terms arise from the Schelling model: if $f_{d}(i) \le T$, the fraction of different neighbors would be under or be equal to the tolerance value, hence the agent would be happy, being $I_{dis}(i)\le 0$. With only these two terms, our model is equivalent to the Schelling one. Nonetheless, we have extended our work considering economic contributions. Mean housing price in the node is denoted as $D(i)$ \cite{Ortega(2021)}. Lower $D(i)$ values imply cheaper living places, thus producing a higher level of happiness. Finally, we have portrayed the economic inequality between groups as $H$, understood as half the economic gap between them. This term favors red agents, $-H$, while blue agents are monetarily handicapped, $+H$. As can be inferred from the previous explanation, lower values of $I_{dis}$ yield a higher level of happiness. To put in in another words, an agent is happy when $I_{dis} \le 0$ and unhappy if $I_{dis} > 0$.

We study an open city framework where agents may abandon the city if they are unhappy, and enter otherwise. Therefore, a large value of $H$ increases the number of red agents in the network while reducing the blue population. In other words, the interplay between $D(i) \pm H$ can generate zones depleted of blue population, which has the bias of the economic gap, $+H$. This term may produce positive values of $I_{dis}(i)$, pointing out the unhappiness of the agent, who can be relocated to a better location if it is available or leave the city. The interpretation is straightforward, the segregated group has no economical access to some city zones, while the red group, which is happier due to the financial advantage, $-H$, resides in them. 

As a means to understand the final state reached, we examine the contribution of each term from Equation~\eqref{eq:1}. The first two terms on the right-hand side give rise to the clustering effect: agents of the same kind minimize these terms by grouping them. If there are no dissimilar agents in the neighborhood, $f_{d}(i)=0$, thus the only effective term is $-T$. An alternative reading of the economic terms is unalike housing prices for red and blue agents: on one hand, living places for the red group can be considered to be priced as $D(i) - H$, while for the economically handicapped blue agents become $D(i) + H$. In other words, prices for blue agents are increased by a factor of $2H$ taking as reference the red group. Consequently, the blue group may only settle in the most affordable living places.

It should be noted that the economic terms can balance the contribution from the neighborhood preferences, as Equation~\eqref{eq:1} points out. For example, agents who are surrounded by neighbors of the same kind may be happy even if the contribution of the housing price and the financial gap create some discomfort, resembling ghettos' reality.

\subsection{System Dynamics}
\label{SD}
We start from a random initial configuration with equal proportions of red and blue agents and a small percentage of vacancies, $5\%$. Nonetheless, the final number of vacancies in the network is controlled by the interplay between $D(i) + H$, so this initial percentage does not vary from our final results. As we consider an open city model, agents can enter or leave the city depending on their dissatisfaction level. It must be noted that $I_{dis}$ are calculated by means of Equation~\eqref{eq:1}.

At each iteration, we choose an internal or external exchange with equal probabilities. On one hand, if the change is internal, two nodes $i$ and $j$ are randomly selected. However, $i$ must be occupied with an agent, while the node $j$ is a vacancy. If the exchange verifies $I_{dis}(j) \leq I_{dis}(i)$ the agent is relocated into the node $j$, and now the node $i$ becomes a vacancy. Otherwise, the relocation is rejected.  On the other hand, if the exchange is external, a random node is selected. If the node $i$ is a vacancy, the node is occupied by an agent coming from outside, if the satisfaction condition is fulfilled, i.e., $I_{dis}(i) \leq 0$. The color of this agent is randomly selected ($50/50$ chance). If the node $i$ has an agent, the agent leaves the system if $I_{dis}(i) > 0$. This iteration is repeated until the system reaches an equilibrium state where all changes are discarded.

In contrast to the Schelling model, happy agents can relocate inside the city if the new place implies a higher level of happiness that the current one. It is also worth mentioning that all the unhappy agents have abandoned the city at the end of the simulation.

\section{Results}
\label{result}

\subsection{Segregation in Washington D.C.}
\label{WDC}
\begin{figure}[H]
\begin{tabular}{cc}
\includegraphics[width=6.00cm, height=4.0cm]{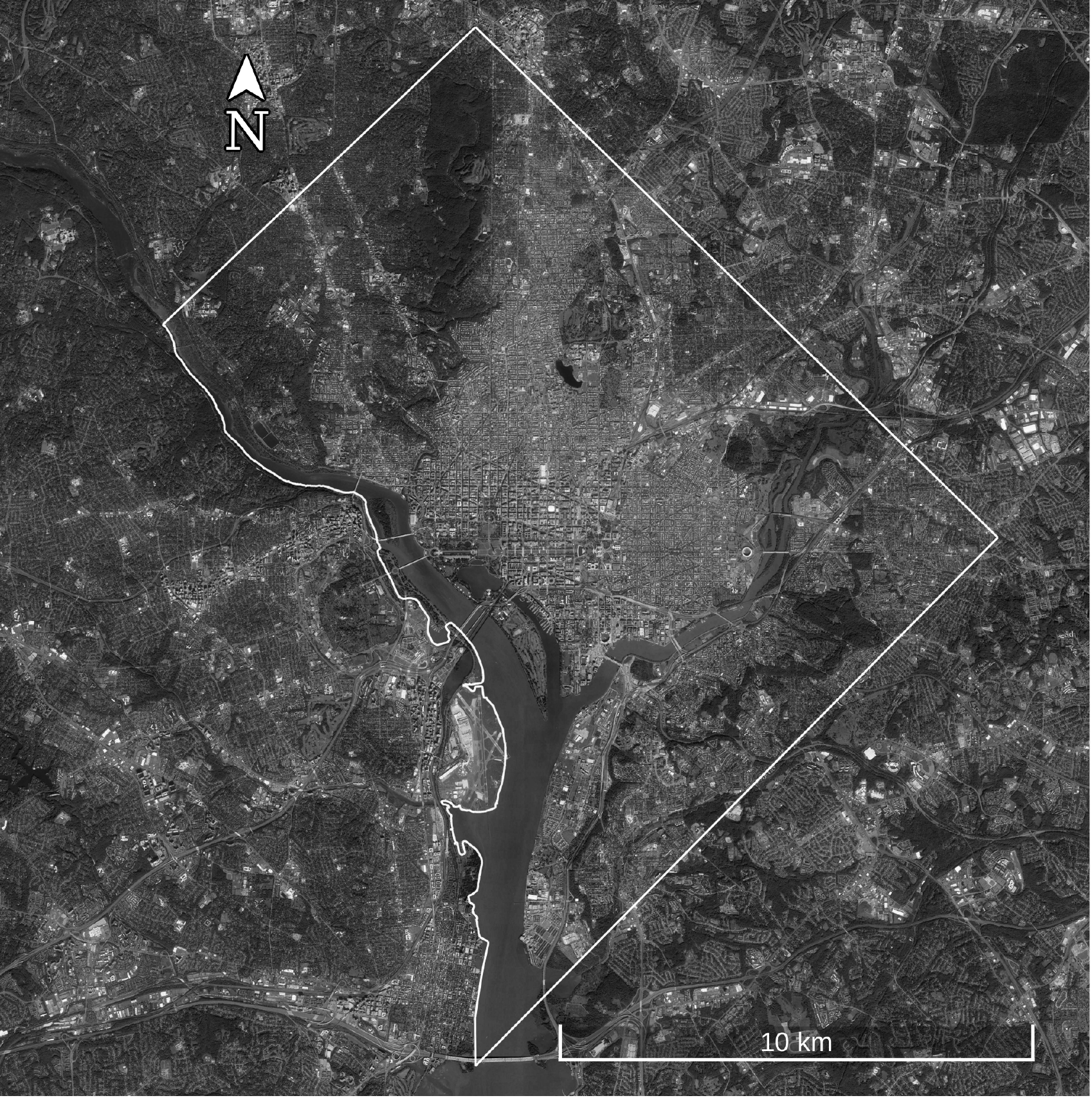}&
\includegraphics[width=6.00cm, height=4.0cm]{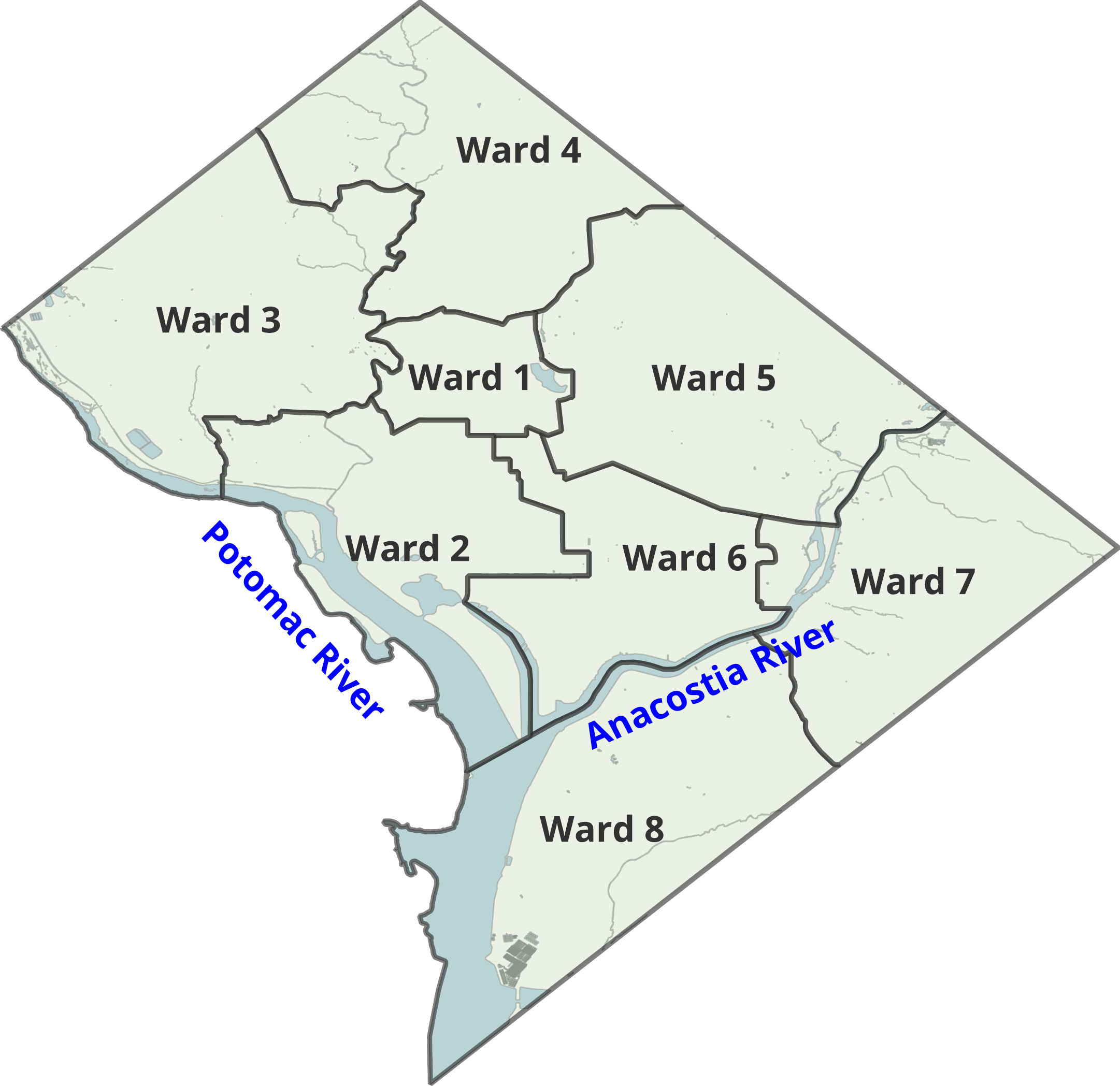}\\
(\textbf{a})&(\textbf{b})\\
\end{tabular}
\caption{(\textbf{a}) Orthoimage of Washington D.C. (\textbf{b}) Division of Washington D.C. by wards. River names are represented in blue.}
\label{fig:3}
\end{figure}

Aiming to study a real case of segregation, we focus our attention on Washington D.C. (USA). This city is located on the eastern shore of the Potomac River, bordering the states of Maryland and Virginia, as we can see in Figure \ref{fig:3}a. The establishment of black ghettos in this city has a historic origin. During the 1960s the city became majority Black and this population was concentrated in the south of Anacostia River, which is depicted in \mbox{Figure \ref{fig:3}b.}  Wards 7 and 8 suffered from disinvestment, and public housing was located in these zones \cite{Asch(2017)}. Consequently, we expect a large concentration of ghettos in this area, which has been segregated for more than 50 years. Although there is no official classification of ghettos, deprived communities such as Anacostia or Deanwood can be currently found in this zone \cite{Areavibes(2022)}.

\subsection{Model}
\label{MR}
Now, we apply the method from Section \ref{GN} to Washington D.C. (USA) considering its division by census tracts. Census tracts are small, relatively permanent statistical subdivisions of a county or statistically equivalent entities with a population size between $1200$ and $8000$ people \cite{Census(2022)}. Data for each tract can be accessed via the $tidycensus$ R \mbox{package \cite{Walker(2022)}.}  After the process, we obtain a network with $179$ nodes and $535$ edges. Additional information on each census tract may be introduced into each node.

\begin{figure}[H]
\begin{tabular}{cc}
\includegraphics[width=6.00cm, height=5cm]{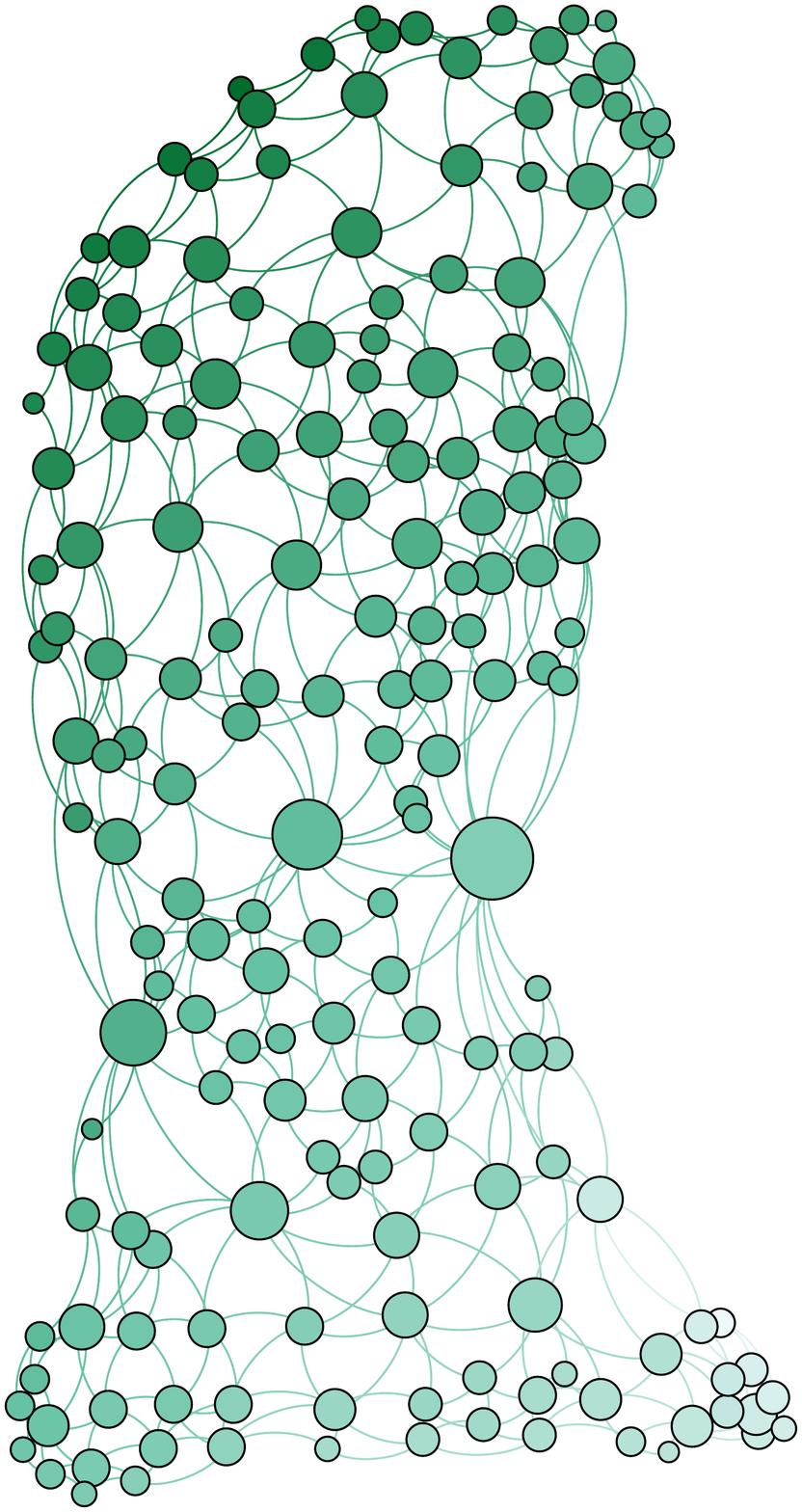}&
\includegraphics[width=6.00cm, height=5cm]{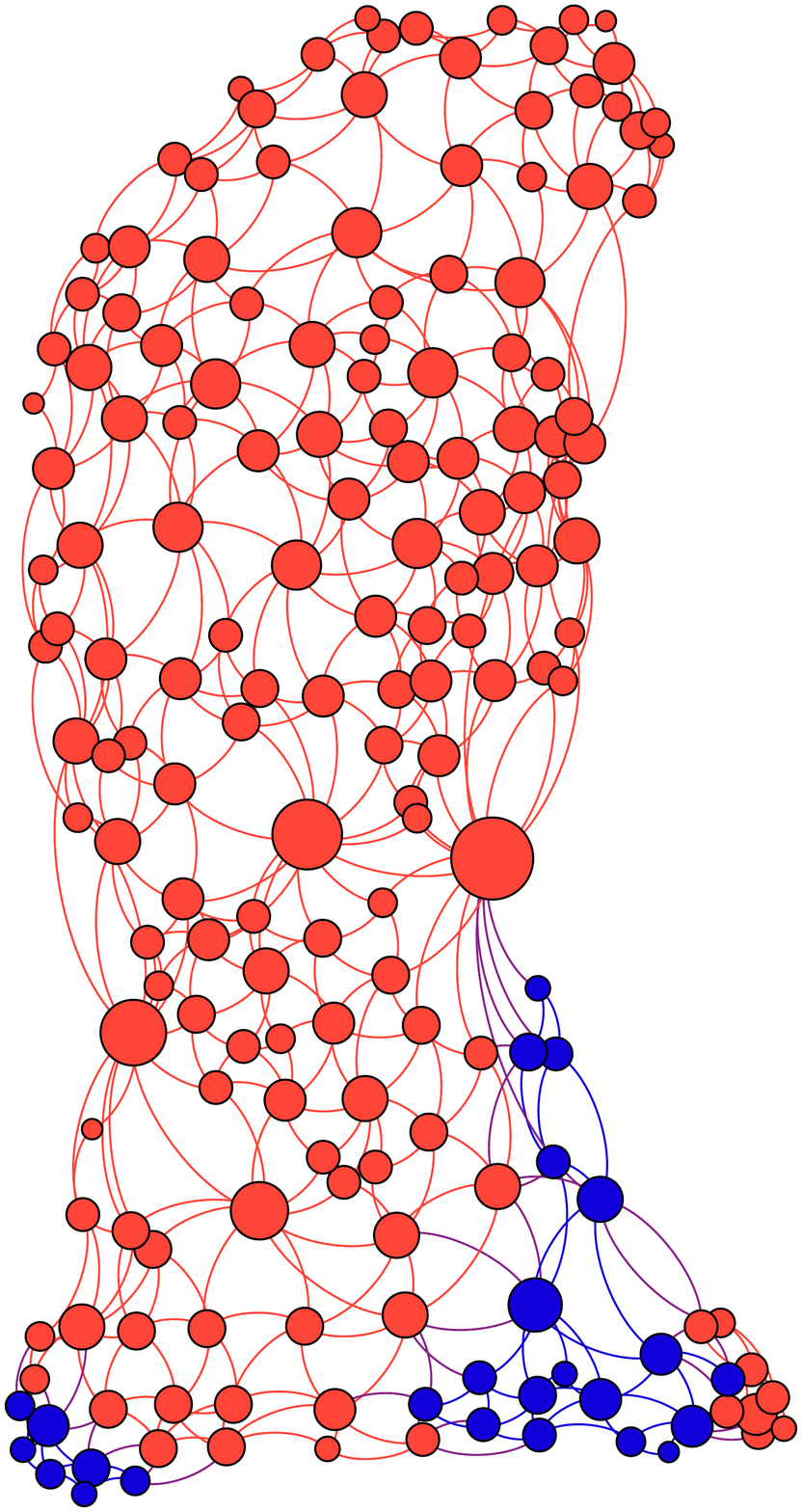}\\
(\textbf{a})&(\textbf{b})\\
\end{tabular}
\caption{(\textbf{a}) Network representation of the housing price $D(i)$ for each node. Darker tones are related to greater values and vice versa. All values are in the range $[-1,0]$. (\textbf{b}) Snapshot of the system's final state for one run. Red and blue agents are represented by their respective colors.}
\label{fig:4}
\end{figure}

We choose the parameters values for Equation~\eqref{eq:1}: $T$, $D(i)$ and $H(i)$. $T$ is a measure of tolerance. In a classic Schelling model, where each agent has 8 neighbors, a tolerance of $T=0.25$ implies that the agent is happy if only two or fewer individuals are different. As it can be seen in Figure 
 \ref{fig:5}b, there are tracts where the white population represents a really high or a very low part of the population. Therefore, we choose $T=0.25$, a low value of tolerance that may explain these extreme racial concentrations. Even though no direct measure of the housing market can be obtained from the database, we can infer it from the median house income Figure \ref{fig:5}a.  This variable seems to decrease from the top to the bottom. Thus, it would be possible to define a vertical gradient for $D(i)$. Extreme cases are $0$, assigned to the census tract with the highest latitude, and $-1$, for the lowest one. The rest of the values lie in the range $[-1,0]$ and are proportional to each centroid latitude. This housing market distribution underlines that the most expensive living places are in the north while the most affordable regions are in the south.  Then, the information is mapped into the network assigning a $D(i)$ value for each node, as can be seen in Figure \ref{fig:4}a. Finally,  $H(i)=0.25$ implies a large financial gap as it can be deduced from the scale of Figure \ref{fig:5}a.

As it is illustrated in Figure \ref{fig:4}b, blue nodes that represent ghettos tend to coincide with the most affordable areas of the city, depicted as clearer nodes in Figure \ref{fig:4}a. The social interpretation is straightforward, segregation also occurs in the economic aspect and people from ghettos do not have the monetary resources to relocate to another place inside the city.
\subsection{Spatial Analysis and Machine Learning Results}
\label{ML}
\begin{figure}[H]
\begin{adjustwidth}{-5cm}{-1.5cm}
\centering
\begin{tabular}{ccc}
\includegraphics[width=5.50cm, height=3.8cm]{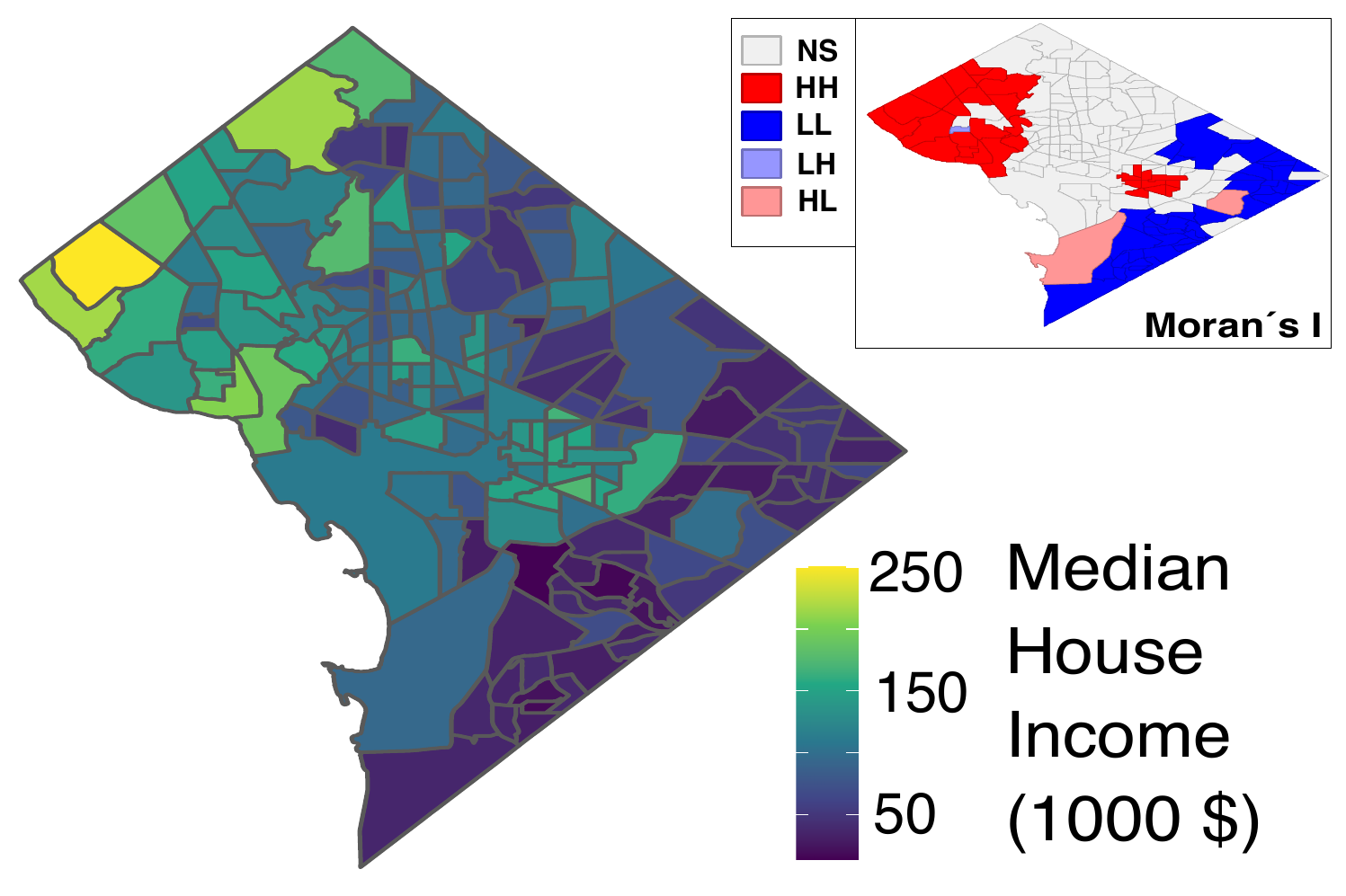}&
\includegraphics[width=5.50cm, height=3.8cm]{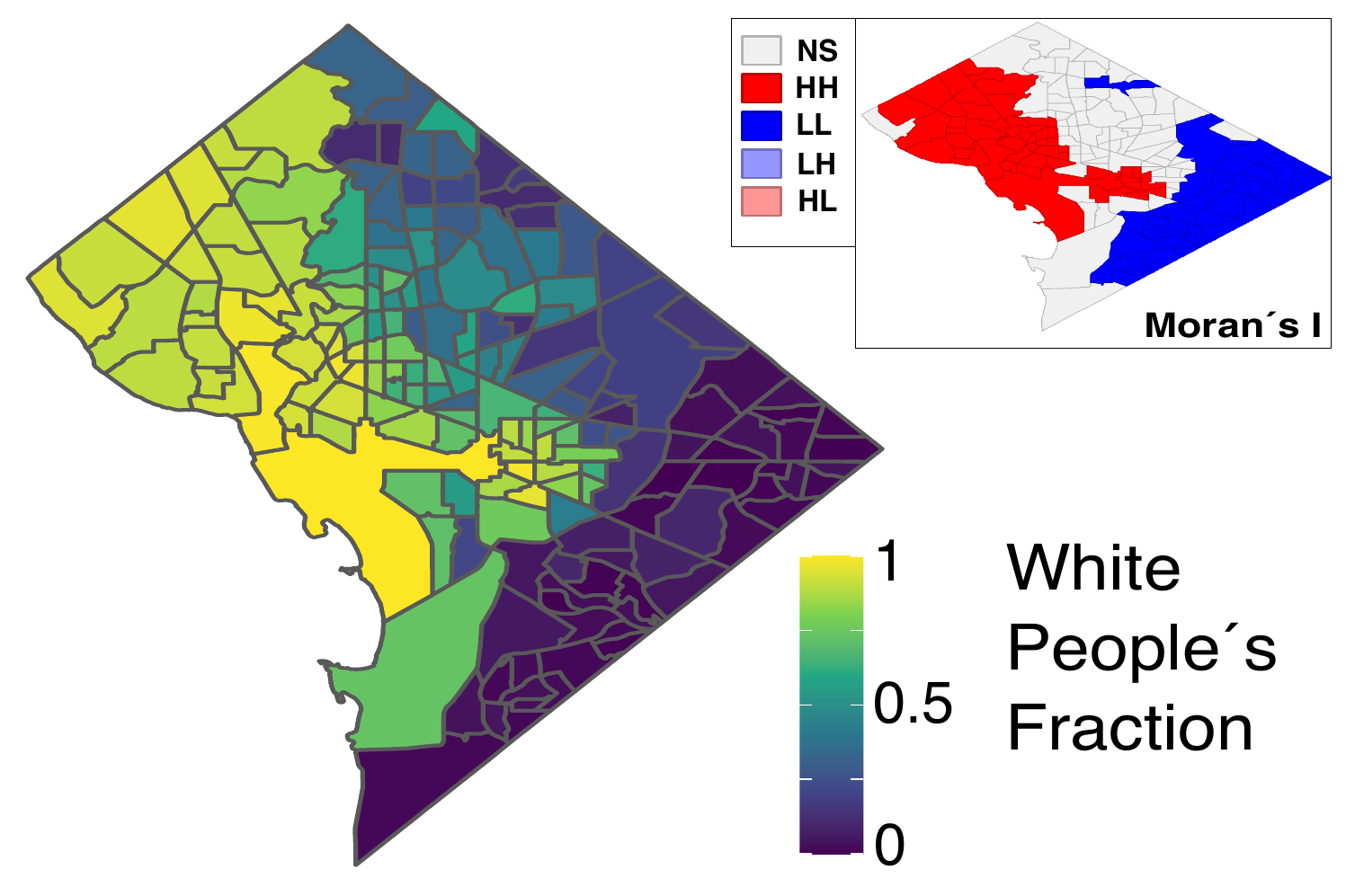}&
\includegraphics[width=5.50cm, height=3.8cm]{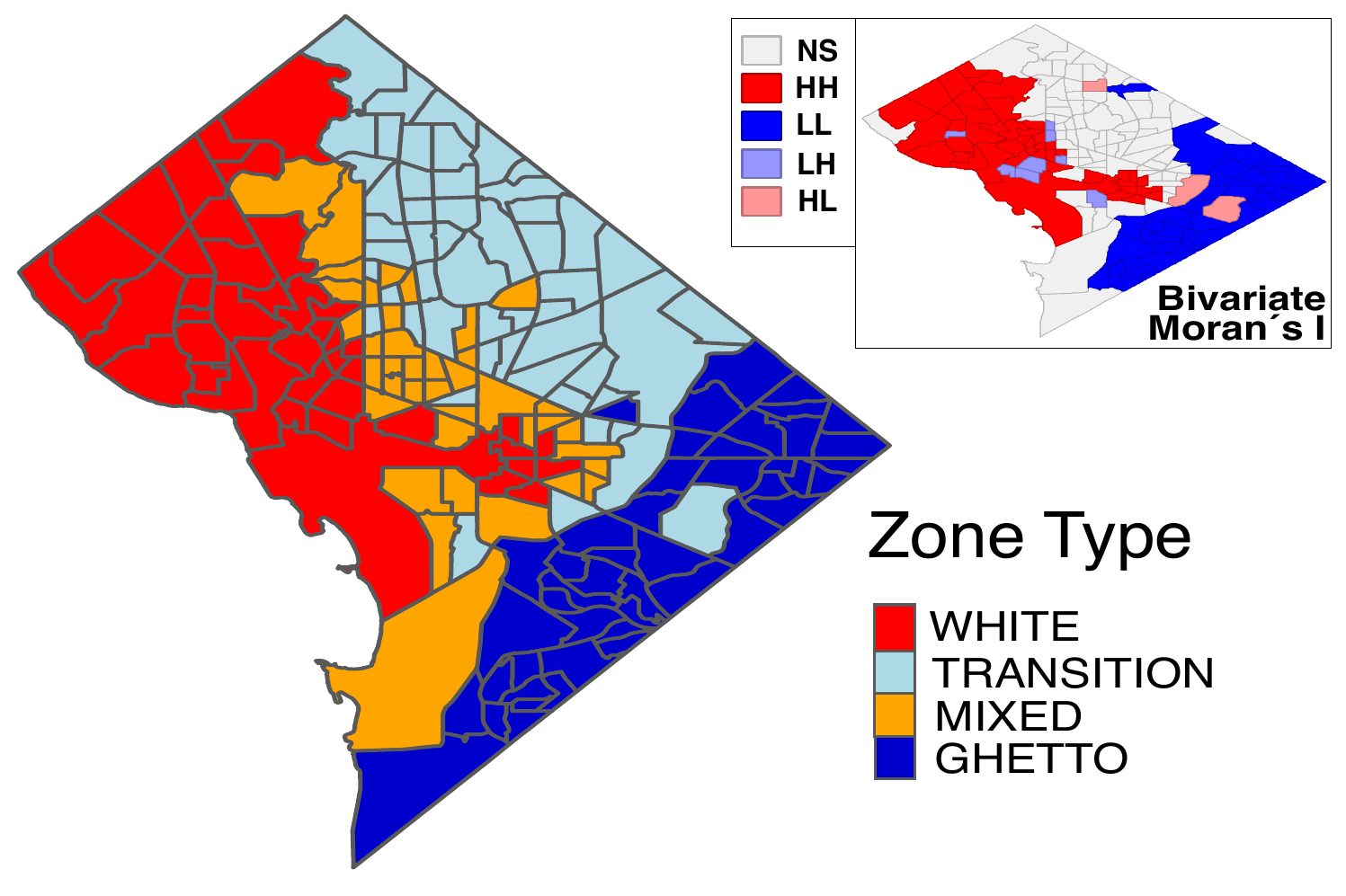}\\
(\textbf{a})&(\textbf{b})&(\textbf{c})\\
\end{tabular}
\end{adjustwidth}
\caption{(\textbf{a}) Median house income distribution for Washington D.C. (\textbf{b}) White people's fraction for our region of interest.  (\textbf{c}) Zone types found by the EM clusterization algorithm. The insets in (\textbf{a}) and (\textbf{b}) represent the local Moran's I.  The inset from (\textbf{c}) illustrates the local Moran's bivariate index choosing as variables the median house income and the white people's fraction. In these insets, NS means Not Significant, while H and L denote High and Low values, respectively. Spatial Analysis software Geoda was used for the insets \cite{Anselin(2006)}.}
\label{fig:5}
\end{figure}

Although segregation is a multiscale phenomenon, we have selected data from the American Community Survey \cite{Walker(2022)} for two of its main expressions: racial and economic. A measure of the economic level is the median house income, which is depicted in Figure \ref{fig:5}a. As can be seen in the figure, the upper part exhibits brighter colors, associated with higher incomes, than the lower ones. Thus, economically handicapped people will be located near the bottom as a consequence of their financial situation, given that the housing market is less expensive in these zones. To study the racial distribution, we calculate the white people's fraction, defined from now on as $WPF$, from the quotient $n_{w}/(n_{w}+n_{b})$  where $n$ denotes the population in the census tract corresponding to white (w) and black (b) races. This coefficient is illustrated in Figure \ref{fig:5}b and follows a similar color scheme to the median house income map from Figure \ref{fig:5}a. Brighter tones are in the upper part of the town, especially concentrated in the west, where the fraction of white people is close to one. In contrast, places for people with financial issues are occupied by a high percentage of black people. These locations can be found in the southeast part of the city and correspond to wards 7 and 8, previously depicted in Figure \ref{fig:3}b.

Economical segregation is estimated by the local Moran’s I in each census tract, as can be seen in the inset of Figure \ref{fig:5}a.  This index is understood as how the value of a variable in a tract relates to the same value in its surroundings \cite{Anselin(1995)}. Thus, the red color in the inset highlights the clustering of rich zones, mainly located in the northwest zone, whereas blue regions in the southeast indicate the grouping of tracts with a handicapped economy. The same analysis for the $WPF$ is graphed in the inset of  Figure \ref{fig:5}b. Results from both insets Figure  \ref{fig:5}a,b exhibit similar patterns. Hence, to test the spatial correlation between economy and race, we use the bivariate local moran's I \cite{Wang(2022)}, which is depicted in the inset of Figure \ref{fig:5}c. This image exhibits a similar pattern to the other insets pointing out the strong correlation between both variables.

In order to identify which census tracts are ghettos, we use the Expectation-Maximization (EM) clusterization algorithm \cite{Batzoglou(2008)}. The EM method is a generalization of the maximum likelihood estimation to the incomplete data case. In particular, the EM algorithm tries to obtain the parameters that maximize the logarithmic probability of the observed data. This method is implemented in Weka, a free ML software \cite{Hall(2009)}. We call this method from the R environment \cite{R(2013)} by using Rweka \cite{Hornik(2009)}, an interface that allows us to run it.  Data supplied to this algorithm are the median house income, the $WPF$, and the latitude of the centroids for each census tract. These data are normalized from 0 to 1. 

The EM method finds four clusters that we have defined as zone types, as can be observed in Figure \ref{fig:5}c. The zone types labeled as \textit{white} and \textit{ghettos} describe regions with opposite situations: high $WPF$ with abundant financial resources and ghettos populated by black people with economical issues. Between them, the \textit{mixed} zone describes a region where the $WPF$ and the median house income have increased a little taking as reference \textit{ghetto} zones. The $transition$ zone is similar to the white zone but the high $WPF$ and income have declined, thus, it can be considered a transition layer from \textit{white} to \textit{mixed} or \textit{ghetto} areas.
Once some tracts are identified as ghettos, we study which kind of segregation is mainly responsible for these areas. Expressed in another way, we try to know if a simple classification for these zone types exists. As a mean, we make use of the J48 algorithm from Rweka \cite{Hornik(2009)}, which is the code name assigned to the ID3 tree classifier \cite{Quinlan(1986)}. This algorithm begins with the original dataset as the root node. On each iteration of the algorithm, it selects the attribute with the largest information gain value, thus creating a new node. Their result is 176 instances correctly classified out of 179. The only variable that the classifier retains is the $WPF$, standing out its importance. Ghettos are identified as tracts where the value of this variable is under $0.098$. 

\subsection{Comparison}
\label{RC}
\begin{figure}[H]
\centering
\includegraphics[width=10.5 cm, height=6cm]{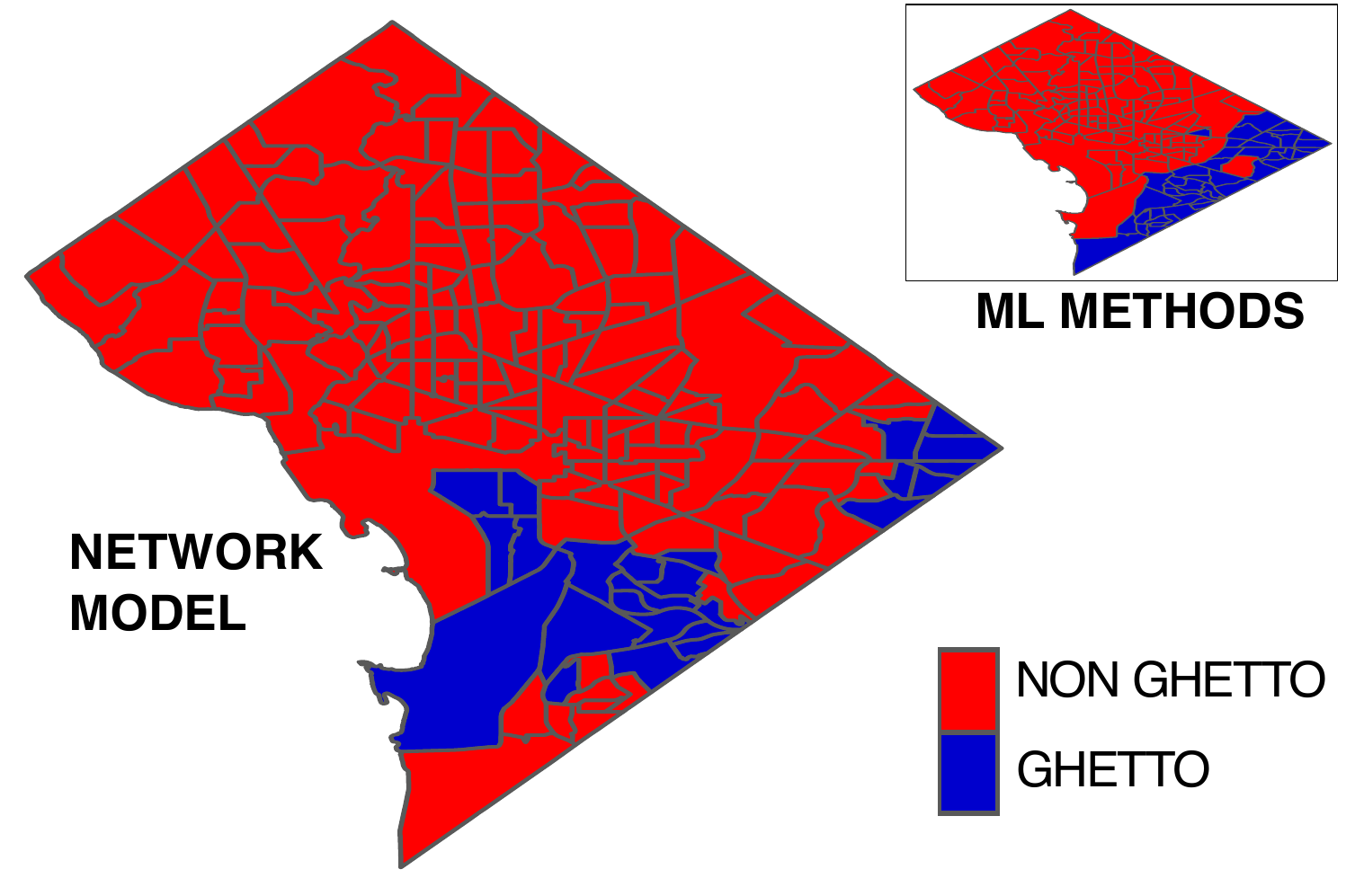}
\caption{Mapping from the network model snapshot (Figure \ref{fig:4}b) to the real city. In order to compare these results with the ones from ML, illustrated in Figure \ref{fig:5}c, the inset exhibit the same zone types.}
\label{fig:6}
\end{figure}

As can be seen in the previous sections, we have obtained the ghettos' location in two ways: a network model and ML methods. First, we compare these results from a qualitative point of view, as it is depicted in Figure \ref{fig:6}. Nodes from the network are identified with census tracts, thus allowing us to map our results into the real city. Classification procedures from the EM algorithm found four types of zones. These sectors are labeled as ghettos and non-ghetto census tracts. It must be noted the resemblance between the model and the ML method, given that both of them locate ghettos in the south region of the city.

We have also evaluated the accuracy of the model, taking the outcome of the ML method as correct. The accuracy is defined as the quotient $R/NT$ where $R$ is the number of census tracts correctly identified and $NT$ is the number of total census tracts, being categorized into the binary classification previously explained. We have simulated 1000 runs finding a mean accuracy of $80 \%$ with a standard deviation of  $7\%$. 

As a means to evaluate the obtained results from this novel approach, it is also interesting to compare our contribution with related works from Section~\ref{lr}. In \cite{Wong(2004)}, Washington D.C. is studied at various levels: 27 zip codes, 188 census tracts, and 433 block groups. Although the main aim of this paper was to study the effects of the modifiable areal unit problem (MAUP), and the data correspond to twenty years ago, we can observe that the proportion of the non-white population is high in the southern part of the city, showing good agreement with our results. In \cite{Benenson(2009)} they applied the Schelling model at the resolution of individual buildings and families to study the ethnic segregation of an area of Tel Aviv. The paper demonstrated good qualitative agreement between the Schelling model and real urban segregation for a period of 40 years. Economic terms were not considered for this study, highlighting the importance of the ethnic group as a basis for segregation. Recently, considering the multiple scales at which segregation occurs, a multilevel index of dissimilarity was defined in \cite{Harris(2017)}. This index was applied to study the residential segregation of various ethnic groups in England and Wales. The results were consistent with a process whereby minority groups had spread out into more mixed neighborhoods. However, the defined index is a general measure of segregation, a general value that can hide spatial heterogeneity.
In \cite{Reitano(2020)} two indices are used: the local spatial dissimilarity and the exposure index. The region of interest was a neighborhood in Naples (Italy). This contribution highlighted the major role played by some policy decisions to reinforce segregation, as it happened in Washington D.C. Another interesting contribution comes from the transportation geography field \cite{Farber(2015)}. The method they put forward decomposed the social interaction potential into interactions within and between social groups. Therefore, they related this potential to segregation by a different race. Data were analyzed at the census tracts level, while a matrix disaggregation technique allow them to obtain the transport fluxes by race. The study allows the identification of hotspots of segregation and integration for the case study of Detroit. In our work, the segregation hotspots are located via spatial analysis.

\section{Discussion}
\label{disc}
In order to understand the segregation in Washington D.C., we begin the discussion with spatial analysis. Moran's indexes from the insets in Figure \ref{fig:5}a,b pinpoint a clustering of white and rich people in the northwest, while the southeast is strongly segregated, showing opposite characteristics. Besides, the bivariate index suggests a strong spatial correlation between the median house income and the white people's fraction, as it is depicted in the inset of Figure \ref{fig:5}c. 

Another issue inherent to SA is the MAUP. Basically, as census tracts or other divisions of the regions can change their boundaries through time, data would have an inherent error. For the Washington D.C. case, the problem was analyzed by \cite{Wong(2004)}. The evaluation of the non-white fraction to the level of entire D.C, tracts, block groups, and ZIP codes gives values of 0.692, 0.724, 0.718, and 0.561, in the previous order. All the estimations are near except the one for the zip code. Therefore, the ZIP code has a large error, being the other measures similar. Thus as we choose the census tracts divisions and the error of our Schelling extended model is larger, the MAUP does not seem an issue in our case. In addition to this, census tract divisions of wards 7 and 8 have a river as a boundary with the rest of the city, as can be observed in Figure \ref{fig:3}b. Hence, this natural boundary which is used for the census tracts has not been modified, thus reducing the MAUP even further. 

Once the spatial analysis was carried out, we studied the data related to the median house income, $WPF$, and latitude from the ML perspective. The EM clusterer found four zone types, as it is shown in Figure \ref{fig:5}c. These results were in good agreement with our SA, and they were taken as the correct classification of the census tracts. In addition to this, a tree classifier ID3 found that the main variable in the previous classification was the $WPF$, meaning that race plays a key role in the zone type. This suggests that the main motive of segregation is racial, even nowadays. However, this racial segregation is strongly linked with economic inequalities, as the SA analysis demonstrated.

Finally, the model results are compared with those from EM clustering reduced to the binary classification of ghettos and non-ghettos, as it is depicted in Figure \ref{fig:6}. Good agreement is found between our model and the ML predictions, obtaining an accuracy of $80 \pm 7\%$. One possible explanation is that the interplay between $D \pm H$ limits the zones where blue clusters can be found to the south of the city. Therefore, it portrays the real situation where ghettos are on the southern shore of the Anacostia river.

\section{Conclusions}
\label{conc}

The extended Schelling model in networks provides a new framework for the study and understanding of ghettos' establishment and their location. Starting with some basic data information from the GIS system, such as the neighborhood relationships between census tracts and a vertical gradient in the house pricing, our network, and its properties are defined. Then, an extended Schelling model, including the economic terms model, runs on it, allowing us to identify which census tracts can be classified as ghettos. It must be noted that this approximation is useful when microdata or reliable data are not available for the zone. As an example, we must mention that due to the impact of the COVID-19 pandemic, the Census Bureau changed the 2020 American Community Survey release into a series of experimental estimates, instead of the standard one \cite{Census(2022)}.

The strong segregation is a consequence of the policies adopted during the 1960s, when the black population was concentrated in the south of Anacostia River (see Figure \ref{fig:3}b),  as was discussed in Section~\ref{WDC}. Then, the disinvestment in the zone caused the population to fall into the poverty trap. This term alludes to self-reinforcing mechanisms that cause poverty to persist unless there is outside intervention \cite{Azariadis(2005)}.
In fact, as we discussed in Section~\ref{DI}, we can consider different prices for the same spot depending on group membership. This fact resembles redlining practices which can be summarized as an increase in the interest rate or even credit denial of a loan due to cultural or racial bias \cite{Rothstein(2018)}. This procedure gave rise to an even further concentration of the deprived black population in the ghetto zones. In addition to this, the river acts as a boundary between this part and the rest of the city which leads to an increase in the zone isolation \cite{Crooks(2009)}. To sum up, the actual setting is strongly influenced by these past practices and the economic gap could act as a \textit{de facto} redlining procedure, not allowing the relocation of economically handicapped people into better locations.

Nevertheless, the model has some limitations: other cities with more complicated structures can create a complex housing market difficult to define. For instance, cities in Europe tend to be radially structured, i.e., they have expanded from the center towards the outskirts where ghettos are mainly located.

A way to enhance this work is to include three parameters in the dissatisfaction index. One is linked to both segregation terms, and the others are associated with the housing market and half the financial gap. In this case, we could maximize the final accuracy of the model by using an optimization procedure over the parameters previously included.

\vspace{6pt}

\authorcontributions{Conceptualization, D.O. and E.K.; methodology, D.O.; software, D.O.; validation, D.O and E.K.; data curation, D.O.; writing---original draft preparation, D.O.; writing---review and editing, E.K.; visualization, D.O.; supervision, E.K.; project administration, E.K.. All authors have read and agreed to the published version of the manuscript.}

\funding{This research was funded by the Spanish Government through grants PGC2018-094763-B-I00 and PID2019-105182GB-I00.}

\institutionalreview{Not applicable.}

\informedconsent{Not applicable.}

\dataavailability{Data were obtained from the American Community Survey (2014-2019) via tidycensus \cite{Walker(2022)}. Data from the median house income, and white and black populations correspond to $B19013\_001$, $C02003\_003$, and $C02003\_004$ variables, respectively. White people refer to persons whose race has a European origin, while black denotes African American people exclusively. Orthophoto images from Washington D.C. in Figures \ref{fig:2}a and \ref{fig:3}a are taken from Google Earth.}

\acknowledgments{We acknowledge financial support from the Spanish Government through grants PGC2018-094763-B-I00 and PID2019-105182GB-I00.}

\conflictsofinterest{The authors declare no conflict of interest.}

\abbreviations{Abbreviations}{
The following abbreviations are used in this manuscript:\\

\noindent 
\begin{tabular}{@{}ll}
BEG & Blume-Emery-Griffiths \\
EM & Expectation-Maximization\\
GIS & Geographic Information Systems\\
ML & Machine Learning\\
MAUP & Modifiable Areal Unit Problem \\
USA & United States of America\\
SA & Spatial Analysis\\
WPF & White People's Fraction
\end{tabular}
}

\appendixtitles{no} %

\end{paracol}
\reftitle{References}

\end{document}